\documentclass[preprint,showpacs,showkeys,preprintnumbers,amsmath,amssymb]{revtex4-1}

\usepackage{graphicx}
\usepackage{dcolumn}
\usepackage{bm}
\newtheorem{theorem}{Theorem}

\begin{document}

\title{BBGKY HIERARCHY AND DYNAMICS OF CORRELATIONS}%

\author{D.O. POLISHCHUK}%

\email{polishuk.denis@gmail.com}

\affiliation{Taras Shevchenko National University of Kyiv,\\
Department of Mechanics and Mathematics
}%

\pacs{05.30.-d, 05.20.Dd, 42.50.Lc, 02.30.Jr, 47.70.Nd.}
\keywords{the von Neumann hierarchy; BBGKY hierarchy; quantum kinetic equations;
cumulant (semi-invariant); cluster expansion; correlation operator; statistical operator
(density matrix); quantum many-particle system; Bose and Fermi statistics}

\begin{abstract}
We derive the BBGKY hierarchy for the Fermi and Bose many-particle systems, using the von Neumann hierarchy for the correlation operators. The solution of the Cauchy problem of the formulated hierarchy for the case of a $n$-body interaction potential is constructed in the space of sequences of trace-class operators.
\end{abstract}

\maketitle
\tableofcontents

\vphantom{math}
\vskip1.5cm

\section{Introduction}
In recent years a large progress in the mathematical theory of the BBGKY hierarchy for quantum many-particle systems is observed. A good example of such progress is the rigorous derivation of quantum kinetic equations that describe Bose condensate \cite{Frol,Mich,ES}.
 In the original works of Bogolyubov \cite{BogGurov,BogQStat,BogBogJr} the solution of the initial-value problem of the BBGKY hierarchy was constructed in the form of the iteration series. The same representation of the solution is also being used in modern works \cite{Petrina70,AA,Got}.

 For the case of the Maxwell-Boltzmann statistics the solution was also constructed in the form of the expansion over particle clusters, which evolution is governed by the cumulants of groups of operators for the von Neumann equations \cite{GerUMJ} (or by reduced cumulants \cite{Petrina}). These solution expansions were constructed on the base of the non-equilibrium grand canonical ensemble \cite{CGP97,Petrina}.

In this paper we propose the alternative method of the description of the evolution of quantum many-particle systems. States of such systems are described in terms of correlation operators, evolution of which is governed by the von Neumann hierarchy. Based on the solution of such hierarchy we define the $s$-particle (marginal) density operators and derive the BBGKY hierarchy \cite{BogGurov}, that can describe the evolution of infinite-particle systems.

Using the solution of the von Neumann hierarchy  for the correlation operators we derive the formula of the solution of the Cauchy problem of the BBGKY hierarchy in the form of the expansion over particle clusters, which evolution is governed by the cumulants of group of operators of finitely many Fermi or Bose particles.

Usual iteration series representation of the solution can be obtained from the constructed solution by the use of analogues of the Duhamel formulas for some classes of interaction potentials.

Let us outline the structure of paper. In section 2 we define the basic notions of the Fermi and Bose many-particle systems and introduce the evolution equations for the correlation operators. In section 3 we introduce the $s$-particle density operators based on the solution of the von Neumann hierarchy for correlation operators and derive the BBGKY hierarchy. In section 4 we construct the solution of the initial-value problem of the BBGKY hierarchy in case of the initial data satisfying the chaos property.

\section{The evolution of correlations of Fermi and Bose many-particle systems}

We consider a quantum system of a non-fixed
(i.e., arbitrary but finite) number of identical (spinless) particles with unit mass $m=1$ in the space $\mathbb{R}^{\nu},$ $\nu\geq 1$
 (in the terminology of statistical mechanics it is known as nonequilibrium grand canonical ensemble \cite{CGP97}), that obey  Fermi-Dirac or Bose-Einstein statistics.

States of the system belong to the space
            $\mathfrak{L}^{1}(\mathcal{F}^{\pm}_\mathcal{H})= \bigoplus_{n=0}^{\infty}
            \mathfrak{L}^{1}(\mathcal{H}^{\pm}_{n})$ of sequences
             $f=\big(f_0,f_{1},\ldots,f_{n},\ldots\big)$ of trace-class operators
            $f_{n}\equiv f_{n}(1,\ldots,n)\in\mathfrak{L}^{1}(\mathcal{H}^{\pm}_{n})$ and $f_0\in \mathbb{C}$,
satisfying the following symmetry condition:
             $f_{n}(1,\ldots,n)=f_{n}(i_{1},\ldots,i_{n}),$
            if  $(i_{1},\ldots,i_{n})\in(1,\ldots,n)$, where $\mathcal{H}$ is a Hilbert space associated with a single particle, $\mathcal{H}^{\pm}_{n}=\mathcal{H}^{{\hat{\otimes} n}}$ is the symmetric (or antisymmetric) tensor product of $n$ Hilbert spaces $\mathcal{H}$;
            $\mathcal{F}^{\pm}_{\mathcal{H}}=\bigoplus_{n=0}^{\infty}\mathcal{H}_{n}^{\pm}$ is the Fock space over the Hilbert space $\mathcal{H}$. Spaces ${\mathfrak{L}^{1} (\mathcal{F}^{\pm}_\mathcal{H})}$
are equipped with the trace norm
\[
            \|f\|_{\mathfrak{L}^{1} (\mathcal{F}^{\pm}_\mathcal{H})} =
            \sum\limits_{n=0}^{\infty} \|f_{n}\|_{\mathfrak{L}^{1}(\mathcal{H}^{\pm}_{n})}
             = \sum\limits_{n=0}^{\infty}~\mathrm{Tr}_{1,\ldots,n}|f_{n}(1,\ldots,n)|.
\]

We denote by $\mathfrak{L}^{1}_{0}$ the everywhere dense set in $\mathfrak{L}^{1}(\mathcal{F}^{\pm}_\mathcal{H})$ of finite sequences
 of degenerate operators \cite{Kato} with infinitely
differentiable kernels with compact supports. Note that the space
$\mathfrak{L}^{1}(\mathcal{F}^{\pm}_\mathcal{H})$ contains sequences of operators more general than those determining
the states of systems of particles. Hereafter we assume that $\mathcal{H}=L^{2}(\mathbb{R}^\nu)$.

The Bose-Einstein and Fermi-Dirac statistics endows the state operators with an additional symmetry properties. We illustrate them on kernels of operators \cite{Petrina}.
Let $f_{n}(q_{1},\ldots,q_{n};$ $q'_{1},\ldots,q'_{n})$ is the kernel of the operator $f_{n}\in\mathfrak{L}^{1}(L^{2,\pm}_{n})$. In case of the Bose-Einstein statistics the kernel is a function that is symmetric  with respect to permutations in each group of arguments:
\[
f_{n}(q_{1},q_{2},...,q_{n};q'_{1},q'_{2},...,q'_{n})= f_{n}(q_{\pi(1)},q_{\pi(2)},...,q_{\pi(n)};q'_{\pi'(1)},q'_{\pi'(2)},...,q'_{\pi'(n)}),
\]

and in case of Fermi-Dirac statistics the corresponding kernel is antisymmetric:
\[
f_{n}(q_{1},q_{2},...,q_{n};q'_{1},q'_{2},...,q'_{n})=(-1)^{(|\pi|+|\pi'|)} f_{n}(q_{\pi(1)},q_{\pi(2)},...,q_{\pi(n)};q'_{\pi'(1)},q'_{\pi'(2)},...,q'_{\pi'(n)}),
\]

where $\pi \in \mathfrak{S}_{n}$ and $\pi' \in \mathfrak{S}_{n}$ are the permutation functions, $\mathfrak{S}_{n}$ is a symmetric group, i.e. the group of all the permutations of the set $(1,2,...,n)$, $|\pi|=0,|\pi'|=0$  if the permutation is even and $|\pi|=1, |\pi'|=1$ if it is odd.

 We define the permutation operator $p_{\pi}:\mathfrak{L}^{1}(\mathcal{H}^{\otimes n})\rightarrow\mathfrak{L}^{1}(\mathcal{H}^{\otimes n})$ in terms of kernels of operators by the following formula:
\[
    (p_{\pi}f_{n})(q_{1},q_{2},...,q_{n};q'_{1},q'_{2},...,q'_{n})=f_{n}(t,q_{\pi(1)},q_{\pi(2)},...,q_{\pi(n)};q'_{1},q'_{2},...,q'_{n}).
\]

Let us introduce the symmetrization operator $\mathcal{S}^{+}_{n} :  \mathfrak{L}^{1}(\mathcal{H}^{\otimes n})\rightarrow\mathfrak{L}^{1}(\mathcal{H}_{n}^{+})$ and the antisymmetrization operator $\mathcal{S}^{-}_{n} : \mathfrak{L}^{1}(\mathcal{H}^{\otimes n})\rightarrow\mathfrak{L}^{1}(\mathcal{H}_{n}^{-})$ by the following formulas:
\begin{equation}\label{Sn}
    \mathcal{S}^{+}_{n}=\frac {1}{n!}\sum\limits_{\pi\epsilon \mathfrak{S}_{n}}p_{\pi},
\end{equation}
\[
\mathcal{S}^{-}_{n}=\frac {1}{n!}\sum\limits_{\pi\epsilon \mathfrak{S}_{n}}(-1)^{|\pi|}p_{\pi}.
\]

The Hamiltonian of the system $H=\bigoplus_{n=0}^{\infty}H_{n}$ is a self-adjoint operator with the domain
$\mathcal{D}(H)=\{\psi=\oplus\psi_{n}\in{\mathcal{F}^{\pm}_{\mathcal{H}}}\mid \psi_{n}\in\mathcal{D}
(H_{n})\in\mathcal{H}^{\pm}_{n},\sum\limits_{n}\|H_{n}\psi_{n}\|^{2}<\infty\}\subset{\mathcal{F}^{\pm}_{\mathcal{H}}}$. On the subspace of infinitely
differentiable symmetric (or antisymmetric) functions with compact supports $\psi_n\in L^{2,\pm}_0(\mathbb{R}^{\nu n})\subset L^{2,\pm}(\mathbb{R}^{\nu n})$
$n$-particle Hamiltonian $H_{n}$ acts according to the formula ($H_{0}=0$)

\begin{equation}\label{Hamiltonian}
H_{n}\psi_n = -\frac{\hbar^{2}}{2}
               \sum\limits_{i=1}^{n}\Delta_{q_i}\psi_n
               +\sum\limits_{k=1}^{n}\sum\limits_{i_{1}<\ldots<i_{k}=1}^{n}\Phi^{(k)}(q_{i_{1}},\ldots,q_{i_{k}})\psi_{n},
\end{equation}
where  $\Phi^{(k)}$ is a $k$-body interaction potential satisfying Kato conditions
 \cite{Kato},  $h={2\pi\hbar}$ is the Planck constant.

 We describe states of the system with Hamiltonian (\ref{Hamiltonian}) by the sequence $g(t)=(0,g_{1}(t,1),\ldots,$ $g_{n}(t,1,\ldots,n),\ldots) \in \mathfrak{L}^{1} (\mathcal{F}^{\pm}_\mathcal{H})$ of correlation operators, which evolution is determined by the initial-value
problem of the von Neumann hierarchy \cite{GerPolArx}:
\begin{equation}\label{vonNeumann}
    \frac{d}{dt}g_{n}(t,Y)=-\mathcal{N}_{n}(Y)g_{n}(t,Y)+
\end{equation}
\[  +\sum\limits_{\mbox{\scriptsize $\begin{array}{c}
       \mathrm{P}:\,Y=\bigcup_{i}X_{i}\,\\|\mathrm{P}|\neq1
       \end{array}$}}\sum\limits_{\mbox{\scriptsize
        $\begin{array}{c}
       {Z_{1}\subset X_{1}},\\Z_{1}\neq\emptyset
       \end{array}$}}\ldots
        \sum\limits_{\mbox{\scriptsize
        $\begin{array}{c}
        {Z_{|\mathrm{P}|}\subset X_{|\mathrm{P}|}},\\Z_{|\mathrm{P}|}\neq\emptyset
        \end{array}$}}(-\mathcal{N}_{int}^{(\sum\limits_{r=1}^{|\mathrm{P}|}|Z_{{r}}|)}
       (Z_{{1}},\ldots,Z_{{|\mathrm{P}|}}))\mathcal{S}^{\pm}_{n}\prod_{X_{i}\subset \mathrm{P}}g_{|X_{i}|}(t,X_{i}),
\]
\begin{equation}\label{vonNeumannInitData}
        g_{n}(t,Y)\big|_{t=0}=g_{n}(0,Y),\qquad n\geq1,
\end{equation}
where $\sum_\mathrm{P} $
    is the sum over all possible partitions of the set $Y=(1,\ldots,n)$ into
    $|\mathrm{P}|$ nonempty mutually disjoint subsets $ X_i\subset Y$, $\sum_{Z_{j} \subset X_{j}}$
 is a sum over all subsets
$Z_{j}\subset X_{j}$, for $f_n\in \mathfrak{L}^{1}_{0}(\mathcal{H}_n^{\pm})\subset\mathcal{D}(\mathcal{N}_n)\subset \mathfrak{L}^{1}(\mathcal{H}_n^{\pm})$
the von Neumann operator $\mathcal{N}_n$ is defined by
\[
   \mathcal{N}_nf_n= -\frac{i}{\hbar}\big(f_{n}H_{n}-H_{n}f_{n}\big),
\]
and
\[
\mathcal{N}^{(k)}_{int}f_n =-\frac{i}{\hbar} (f_n\Phi^{(k)}-\Phi^{(k)}f_n).
\]
In case of the Maxwell-Boltzmann statistics this hierarchy was studied in the work \cite{JStatMech}.

We note that the relation between correlation operators defined by hierarchy (\ref{vonNeumann})-(\ref{vonNeumannInitData}) and density operators $D(t)\in \mathfrak{L}^{1} (\mathcal{F}^{\pm}_\mathcal{H})$ has the following form:

\begin{equation}\label{gtdt}
g_{n}(t,Y)=D_{n}(t,Y)+\sum\limits_{\mbox{\scriptsize $\begin{array}{c}
       \mathrm{P}:\,Y=\bigcup_{i}X_{i},\\|\mathrm{P}|\neq1
       \end{array}$}}
       (-1)^{|\mathrm{P}|-1} (|\mathrm{P}| -1)! \,\,{\mathcal{S}^{\pm}_{n}}\prod_{X_i\subset \mathrm{P}}D_{|X_i|}(t,X_i),
\end{equation}
where the sequence $D(t)$ is the solution of the Cauchy problem of the von Neumann equation \cite{Petrina}.

Further we consider a more general notion, namely, the correlation operators of particle clusters that describe the correlations between clusters of particles.

We introduce the following notations: $Y_\mathrm{P}\equiv(\{X_1\},\ldots,\{X_{|\mathrm{P}|}\})$ is a set, which elements are $|\mathrm{P}|$ mutually disjoint subsets $X_i\subset Y\equiv (1,\ldots,s)$ of the partition $\mathrm{P}:Y=\bigcup_{i=1}^{|\mathrm{P}|}X_i.$ Since $Y_\mathrm{P}\equiv(\{X_1\},\ldots,\{X_{|\mathrm{P}|}\}), \{Y\}$ is the set that consists of one element $Y=(1,\ldots,s)$ of the partition $\mathrm{P } (|\mathrm{P}|=1)$.

We define the declasterization mapping $\theta$ by the following formula:
\[
    \theta(Y_P)=Y.
\]
Let us consider the set $X_c=(\{Y\},s+1,\ldots,s+n)$. It holds:
\[
    \theta(X_c) = X \equiv (1,\ldots,s+n).
\]

The relation between correlation operators of particle clusters $g(t)=(0,g_1(t,\{Y\}),\ldots,$ $g_{1+n}(X_c),\ldots) \in \mathfrak{L}^{1} (\mathcal{F}^{\pm}_\mathcal{H})$ and correlation operators of particles (\ref{gtdt}) is given by:
\begin{equation}\label{gCluster}
g_{1+n}(t,X_c)=
\sum\limits_{\mathrm{P}:X_c=\bigcup_i X_i}
       (-1)^{|\mathrm{P}|-1}(|\mathrm{P}| -1)! \mathcal{S}^{\pm}_{s+n}\prod_{X_i\subset \mathrm{P}}
       \sum\limits_{\mathrm{P'}:\,\theta(X_{i})=\bigcup_{j} Z_{ij}} \prod_{Z_{ij}\subset \mathrm{P'}}g_{|Z_{ij}|}(t,Z_{ij}).
\end{equation}

Using hierarchy (\ref{vonNeumann}) from formula (\ref{gCluster})
we derive the generalized von Neumann hierarchy for the correlation operators of particle clusters
\begin{equation}\label{dgdt}
    \frac{d}{dt}g_{1+n}(t,X_c)
    =-\mathcal{N}_{s+n}(X)g_{1+n}(t,X_c)+
\end{equation}
\[
+\mathcal{S}^{\pm}_{s+n}\sum\limits_{\mbox{\scriptsize $\begin{array}{c}
       {\mathrm{P}}:X_c
       ={\bigcup\limits}_i X_i,\\|\mathrm{P}|>1
       \end{array}$}}
       \sum\limits_{\mbox{\scriptsize $\begin{array}{c}
       Z_1 \subset \theta(X_1),\\|Z_1|\geq1
       \end{array}$}}\ldots\sum\limits_{\mbox{\scriptsize $\begin{array}{c}
       Z_{|\mathrm{P}|} \subset \theta(X_{|\mathrm{P}|}),\\|Z_{|\mathrm{P}|}|\geq 1
       \end{array}$}}(-\mathcal{N}_{int}^{(\sum\limits_{i=1}^{|\mathrm{P}|}|Z_{i}|)}(Z_1,\ldots,Z_{|\mathrm{P}|}))\prod\limits_{X_i \subset \mathrm{P}}g_{|X_i|}(t,X_{i}),
\]
where $X\equiv(1,\ldots,s+n)$.

\section{The derivation of the BBGKY hierarchy from the dynamics of correlations}

Let us introduce the $s$-particle (marginal) density operators using the correlation operators that satisfy hierarchy (\ref{dgdt}) by the following formula:
\begin{equation}\label{Fs}
F_{s}(t,Y):=\sum\limits_{n=0}^{\infty}\,\frac{1}{n!}\,\,\mathrm{Tr}_{s+1,\ldots,s+n}\,g_{1+n}(t,X_c),
\end{equation}
where $Y=(1,\ldots,s)$, $X_c=(\{Y\},s+1,\ldots,s+n)$. Series (\ref{Fs}) is convergent in $\mathfrak{L}^{1}(\mathcal{F}^{\pm}_\mathcal{H})$ if $g_{1+n}(t) \in \mathfrak{L}^{1}(\mathcal{H}^{\pm}_{s+n})$.

We show that evolution of marginal density operators defined by expansion (\ref{Fs}) is governed by the chain of equations introduced by Bogolyubov \cite{BogQStat}. The following derivation is given for the case of a two-body interaction potential, i.e. the terms $\mathcal{N}_{int}^{(l)}$ with $l>2$ are equal to zero.

Let us differentiate both sides of expansion (\ref{Fs}) over the time variable in the sense of pointwise convergence in $\mathfrak{L}^{1}(\mathcal{F}^{\pm}_\mathcal{H})$ and use equality (\ref{dgdt}):
\[
       \frac{d}{dt}F_{s}(t,Y)=\sum\limits_{n=0}^{\infty}\frac{1}{n!}\mathrm{Tr}_{s+1,\ldots,s+n}\Big(
       -\mathcal{N}_{s+n}(X)g_{1+n}(t,X_c)+
\]
\[
        +\mathcal{S}^{\pm}_{s+n}
\sum\limits_{\mbox{\scriptsize $\begin{array}{c}
       {\mathrm{P}}:X_c
       = X_1\bigcup X_2
       \end{array}$}}\sum\limits_{i_1 \in \theta(X_1)}\sum\limits_{i_2 \in \theta(X_2)}(-\mathcal{N}_{int}^{(2)}(i_1,i_2)) g_{|X_1|}(t,X_{1})g_{|X_2|}(t,X_{2})\Big),
\]
where $X=(1,\ldots,s+n)$ and  the operators $\mathcal{S}^{\pm}_{s+n}$ are defined by (\ref{Sn}).
Taking into account the fact that
\[
\mathcal{N}_{s+n}(X)=\mathcal{N}_s(Y)+\mathcal{N}_n(X\setminus Y)+\sum\limits_{i_1\subset Y}\sum\limits_{i_2\subset X\setminus Y}(\mathcal{N}_{int}^{(2)}(i_1,i_2)),
\]
 and that for $g_{1+n}(t) \in \mathfrak{L}^{1}(\mathcal{H}^{\pm}_{s+n})$ it holds
\[
    \mathrm{Tr}_{s+1,\ldots,s+n}(-\mathcal{N}_{n}(X\setminus Y)) g_{1+n}(t,X_c)=0,
\]
we rewrite the last equation in the following form:
\[
\frac{d}{dt}F_{s}(t,Y)=\sum\limits_{n=0}^{\infty}\frac{1}{n!}\mathrm{Tr}_{s+1,\ldots,s+n}\Big(\big(
       -\mathcal{N}_{s}(Y) +\sum\limits_{i_1 \in Y}\sum\limits_{i_2 \in X\backslash Y}(-\mathcal{N}_{int}^{(2)}(i_1,i_2))
\big)g_{1+n}(t,X_c)+
\]
\[
       +\mathcal{S}^{\pm}_{s+n} \sum\limits_{\mbox{\scriptsize $\begin{array}{c}
       {\mathrm{P}}:X_c= X_1\bigcup X_2
       \end{array}$}}
       \sum\limits_{i_1 \in \theta(X_1)}\sum\limits_{i_2 \in \theta(X_2)}(-\mathcal{N}_{int}^{(2)}(i_1,i_2))g_{|X_1|}(t,X_{1})g_{|X_2|}(t,X_{2})\Big).
\]
Since operator $\mathcal{N}_s(Y)$  does not depend on variables $s+1,\ldots,s+n$ and according
to the symmetry properties of operators $g_{1+n}(t)$, we obtain:
\[
\frac{d}{dt}F_{s}(t,Y)=-\mathcal{N}_s(Y)\sum\limits_{n=0}^{\infty}\frac{1}{n!}\mathrm{Tr}_{s+1,\ldots,s+n}g_{1+n}(t,X_c)+
\]
\[
+\sum\limits_{n=0}^{\infty}\frac{1}{n!}\mathrm{Tr}_{s+1,\ldots,s+n}\Big(n \sum\limits_{i \in Y}(-\mathcal{N}_{int}^{(2)}(i,s+1))g_{1+n}(t,X_c)+
\]
\[
+\mathcal{S}^{\pm}_{s+n} \sum\limits_{\mbox{\scriptsize $\begin{array}{c}
       {\mathrm{P}}:X_c
       = X_1\bigcup X_2
       \end{array}$}}\sum\limits_{i_1 \in \theta(X_1)}\sum\limits_{i_2 \in \theta(X_2)}(-\mathcal{N}_{int}^{(2)}(i_1,i_2))g_{|X_1|}(t,X_{1})g_{|X_2|}(t,X_{2})\Big).
\]
According to definition (\ref{Fs}),  first term on the right-hand side of the equation is equal to $(-\mathcal{N}_sF_s)$.
Using the symmetry property of the product $g_{|X_1|}(t,X_1)g_{|X_2|}(t,X_2)$ which is the consequence of being under the trace sign,
we make the following rearrangement in the last term:
\[
\sum\limits_{\mathrm{P}:(X_c,s+n+1)=X_1\cup X_2}\,\sum\limits_{i_1 \in \theta(X_1)}\sum\limits_{i_2 \in \theta(X_2)}(-\mathcal{N}_{int}^{(2)}(i_1,i_2))g_{|X_1|}(t,X_1)g_{|X_2|}(t,X_2)=
\]
\[
=\sum\limits_{k=1}^{n+1}\sum\limits_{\mbox{\scriptsize $\begin{array}{c} Z \subset(s+1,\ldots,s+n+1), \\ |Z|=k \end{array}$}}
\sum\limits_{i_1 \in Y}\sum\limits_{i_2 \in Z}(-\mathcal{N}_{int}^{(2)}(i_1,i_2))g_k(t,Z)g_{2+n-k}(t,(X_c,s+1+n)\backslash Z).
\]

As the result we obtain:
\[
\frac{d}{dt}F_{s}(t,Y)=-\mathcal{N}_s(Y)F_s(Y)+
\]
\[
+\sum\limits_{n=0}^{\infty}\frac{1}{n!}\mathrm{Tr}_{s+1,\ldots,s+n+1}\Big(\sum\limits_{i \in Y}(-\mathcal{N}_{int}^{(2)}(i,s+1))g_{2+n}(t,X_c,s+n+1)+
\]
\[
    +\frac{1}{n+1}\mathcal{S}^{\pm}_{s+n+1}\sum\limits_{k=1}^{n+1}\sum\limits_{
       \mbox{\scriptsize $\begin{array}{c} Z\subset(s+1,\ldots,s+1+n), \\ |Z|=k \end{array}$}}
       \sum\limits_{i_1 \in Y}\sum\limits_{i_2 \in Z} (-\mathcal{N}_{int}^{(2)}(i_1,i_2)) g_k(t,Z)\times\\
\]
\[
       \times g_{2+n-k}(t,(X_c,s+1+n)\backslash Z) \Big).
\]
Then we get trace over $s+1$ index out of the sum and use the symmetry property to rewrite the last term:
\[
\frac{d}{dt}F_{s}(t,Y) = -\mathcal{N}_s(Y)F_s(t,Y) +
\]
\[
+\mathrm{Tr}_{s+1}\sum\limits_{i \in Y} (-\mathcal{N}_{int}^{(2)}(i,s+1)) \sum\limits_{n=0}^{\infty}\frac{1}{n!}\mathrm{Tr}_{s+2,\ldots,s+n+1}\Big(g_{2+n}(t,X_c,s + n + 1)+
\]
\[
+\mathcal{S}^{\pm}_{s+n+1} \sum\limits_{k=1}^{n+1}\frac{n!}{(k-1)!(n+1-k)!} g_k(t,s+n+2-k,\ldots,s+n+1)\times
\]
\[
\times g_{2+n-k}(t,\{Y\},s+1,\ldots,s+n-k+1) \Big).
\]

Now, to finish the derivation we need an auxillary fact. In view of the symmetry of operators $g_{1+n}(t)$ with respect to $s+1,\ldots,s+1+n$ variables under the corresponding trace signs, it holds:
\[
g_{1+n}(t,\{1,\ldots,s+1\},s+2,\ldots,s+n+1)= g_{n+2}(t,\{1,\ldots,s\},s+1,\ldots,s+n+1)+
\]
\[
+\mathcal{S}^{\pm}_{s+n+1} \sum\limits_{k=1}^{n+1} \frac{n!}{(k-1)!(n-k+1)!}g_k(t,s+n+2-k,\ldots,s+n+1)\times
\]
\[
\times g_{2+n-k}(t,\{1,\ldots,s\},s+1,\ldots,s+n-k+1).
\]

Thus, we obtain the following equality:
\[
\frac{d}{dt}F_{s}(t,Y)= -\mathcal{N}_s(Y)F_s(t,Y) +
\]
\[
+   \mathrm{Tr}_{s+1}\sum\limits_{i \in Y} (-\mathcal{N}_{int}^{(2)}(i,s+1))\sum\limits_{n=0}^{\infty}\frac{1}{n!}\mathrm{Tr}_{s+2,\ldots,s+n+1}g_{1+n}(t,\{1,\ldots,s+1\},s+2,\ldots,s+n+1),
\]
and according to definition (\ref{Fs}), we deduce finally:
\[
\frac{d}{dt}F_{s}(t,Y)=-\mathcal{N}_s(Y)F_s(t,Y)+\mathrm{Tr}_{s+1}\sum\limits_{i \in Y} (-\mathcal{N}_{int}^{(2)}(i,s+1))F_{s+1}(t,Y,s+1).
\]
This equality can be treated as hierarchy of evolution equations for the marginal density operators and was derived by Bogolyubov in work \cite{BogQStat} from the von Neumann equation for a system of fixed number of particles. For the case of a $n$-body interaction potential we derive the BBGKY hierarchy in the similar way. Its initial-value problem has the following form
\begin{eqnarray}\label{ddtFs2}
  &&\frac{d}{dt}F_{s}(t,Y)=-\mathcal{N}_{s}(Y)F_{s}(t,Y)+\\
  &&+\sum\limits_{n=1}^{\infty}\frac{1}{n!} \mathrm{Tr}_{s+1,\ldots,s+n} \sum\limits_{\mbox{\scriptsize $\begin{array}{c}
       {Z\subset Y},\\Z\neq\emptyset
       \end{array}$}}
        \Big(-\mathcal{N}_{int}^{(|Z|+n)}(Z,X\backslash Y)\Big)F_{s+n}(t,X),\nonumber
\end{eqnarray}
\begin{eqnarray}\label{ddtFs0}
  &&F_{s}(t)\mid_{t=0}=F_{s}(0),\quad s\geq 1.
\end{eqnarray}


\section{On the solution of the Cauchy problem of the BBGKY hierarchy}
Let us construct the solution of the Cauchy problem of the BBGKY hierarchy (\ref{ddtFs2})-(\ref{ddtFs0}) for one physically motivated example of initial data, namely the case of initial data satisfying the chaos property. A
chaos property means that there are no correlations in the system at the initial moment of time (t=0):
\begin{equation}\label{gChaos}
g(0)=(g_1(0,\{Y\}),0,0,...),
\end{equation}
 which means in terms of marginal density operators (\ref{Fs}) that
\begin{equation}\label{Fsg1}
F_s(0,Y)=g_1(0,\{Y\}),
\end{equation}
where $Y=(1,\ldots,s)$.

Let us denote by $\mathfrak{A}_{1+n}(t,X_c)$ the $(1+n)$-order, $n\geq1$, cumulant of groups of operators
\[
\mathcal{G}_{n}(-t)f_{n} =e^{-{\frac{i}{\hbar}}tH_{n}}f_{n}e^{{\frac{i}{\hbar}}tH_{n}},
\]
which is defined by formula:

\begin{equation}\label{Ut}
    \mathfrak{A}_{1+n}(t,X_c):=\sum\limits_{\mathrm{P}:\,X_{c}=
    \bigcup_kZ_k}
    (-1)^{|\mathrm{P}|-1}({|\mathrm{P}|-1})! \prod\limits_{Z_k\subset\mathrm{P}}\mathcal{G}_{|\theta(Z_{k})|}(-t,\theta(Z_{k})),
\end{equation}
where $H_n$ is the Hamiltonian (\ref{Hamiltonian}), $f_n\in\mathfrak{L}^{1}(\mathcal{H}^{\pm}_n)$, $\theta$ is the declasterization mapping defined in section 2 and $X_c=(\{Y\},s+1,\dots,s+n)$.

In case of the Bose or Fermi system of particles a solution of the initial-value problem of the von Neumann hierarchy (\ref{dgdt}) for initial data (\ref{gChaos}) has the following form
\begin{equation}\label{gtUg0}
g_{1+n}(t,X_c)= \mathfrak{A}_{1+n}(t,X_c)\,\mathcal{S}_{s+n}^{\pm}\,\prod\limits_{i\in X_c}g_{1}(0,i),
\end{equation}
where $\mathfrak{A}_{1+n}(t,X_c)$ is given by (\ref{Ut}) and operators $\mathcal{S}_{s+n}^{\pm}$  are defined by (\ref{Sn}).
According to formula (\ref{gtUg0}), from definition (\ref{Fs}) we obtain
\[
F_s(t,Y)=\sum\limits_{n=0}^{\infty}\,\frac{1}{n!}\,\mathrm{Tr}_{s+1,\ldots,s+n}\,\mathfrak{A}_{1+n}(t,X_c)\,
\mathcal{S}_{s+n}^{\pm} \prod\limits_{i\in X_c}g_1(0,i).
\]
In view of equality (\ref{Fsg1}), i.e.
\[g_1(0,i)=\prod\limits_{j\in \theta(i)}F_1(0,j),\]
finally we obtain
\[
F_s(t,Y)= \sum\limits_{n=0}^{\infty}\,\frac{1}{n!}\,\mathrm{Tr}_{s+1,\ldots,s+n}\,\mathfrak{A}_{1+n}(t,X_c)\,\mathcal{S}_{s+n}^{\pm} \,\prod\limits_{i=1}^{s+n}F_1(0,i),
\]
which is the solution of the Cauchy problem of the BBGKY hierarchy (\ref{ddtFs2})-(\ref{ddtFs0}) for the Fermi or Bose many-particle system with initial data satisfying the chaos property.

For arbitrary initial data $F(0)\in\mathfrak{L}_{\alpha}^1(\mathcal{F}^{\pm}_{\mathcal{H}})=\bigoplus_{n=0}^{\infty}\alpha^n\mathfrak{L}^{1}(\mathcal{H}^{\pm}_n)$,
the following theorem holds.
\begin{theorem}
If $F(0)\in\mathfrak{L}_{\alpha}^1(\mathcal{F}^{\pm}_{\mathcal{H}})=\bigoplus_{n=0}^{\infty}\alpha^n\mathfrak{L}^{1}(\mathcal{H}^{\pm}_n)$,
then for
$\alpha>e$ and $t\in \mathbb{R}$ there exists a unique solution of the initial value problem (\ref{ddtFs2})-(\ref{ddtFs0}) given by series
\begin{equation}\label{FsSolution}
F_s(t,1,\ldots,s)=\sum\limits_{n=0}^{\infty}\,\frac{1}{n!}\,\mathrm{Tr}_{s+1,\ldots,s+n}\,\mathfrak{A}_{1+n}(t,X_c)\,F_{s+n}(0,1,\ldots,s+n).
\end{equation}
For initial data $F(0)\in \mathfrak{L}^1_{\alpha,0}(\mathcal{F}^{\pm}_{\mathcal{H}})\in\mathfrak{L}^1_{\alpha}(\mathcal{F}^{\pm}_{\mathcal{H}})$ expansion (\ref{FsSolution}) is a strong solution and for arbitrary initial data from the space $\mathfrak{L}^1_{\alpha}(\mathcal{F}^{\pm}_{\mathcal{H}})$ it is a weak solution.
\end{theorem}

In case of the Maxwell-Boltzmann statistics a solution of the BBGKY hierarchy (\ref{ddtFs2})-(\ref{ddtFs0}) for initial data from the space of trace-class operators was constructed in \cite{GerUMJ}, \cite{GVI, GVIa}.

\section{Conclusion}
In the paper the marginal density operators were defined by the means of a solution of the von Neumann hierarchy (\ref{dgdt}) for correlation operators by formula (\ref{Fs}). This definition allowed to construct the BBGKY hierarchy for the Fermi and Bose particles on the base of dynamics of correlations in the space of sequences of trace-class operators.

We note that using the definition (\ref{Fs}) of marginal density operators it is possible to justify the BBGKY hierarchy in other Banach spaces that contain the states of infinite-particle systems, contrary to the definition of the marginal density operators in the framework of non-equilibrium grand canonical ensemble \cite{Petrina}.

In the paper it was also defined the solution (\ref{FsSolution}) of the BBGKY hierarchy (\ref{ddtFs2}) for the Fermi and Bose particles with a $n$-body interaction potential. Such solution is represented in the form of the expansion (\ref{FsSolution}) over the clusters of particles, which evolution is governed by the cumulants (\ref{Ut}) of the groups of operators for the von Neumann equations. These cumulants for the Fermi and Bose particles have the same structure, as in the Maxwell-Boltzmann case \cite{GerUMJ}.



\begin{thebibliography}{99}

\bibitem{Frol}      J. Fr\"{o}hlich, S. Graffi and S. Schwarz, Commun. Math. Phys. {\bf 271}, 681  (2007).

\bibitem{ES}        L. Erd\"{o}s, B. Schlein and H.-T. Yau, Invent. Math. {\bf 167}, 515 (2007).

\bibitem{Mich}      A. Michelangeli, e-print s.i.s.s.a. 70/2007/mp (2007).

\bibitem{BogQStat}  N.N. Bogolyubov, \emph{Lectures on Quantum Statistics. Problems of Statistical Mechanics of Quantum Systems}
                                      (Rad. Shkola, Kyiv, 1949) (in Ukrainian).

\bibitem{BogGurov}  N.N. Bogolyubov and K.P. Gurov, J. Exp. Theor. Phys. {\bf 17}, 614 (1947).

\bibitem{BogBogJr}  N.N. Bogolyubov and N.N. Bogolubov, Jr. \emph{Introduction to Quantum Statistical Mechanics},
                                                               (Gordon and Breach, New York, 1992).

\bibitem{Petrina70} D.Ya. Petrina, Theor. Math. Phys. {\bf 4}, 394 (1970).

\bibitem{AA}        A. Arnold, Lect. Notes in Math. {\bf 1946}, 45 (2008).

\bibitem{Got}       A.D. Gottlieb  and N.J. Mauser, Phys. Rev. Let. {\bf 95}, 213 (2005).

\bibitem{GerUMJ}    V.I. Gerasimenko and V.O. Shtyk,  Ukr. Math. J. {\bf 58}, 1175 (2006).

\bibitem{Petrina}   D.Ya. Petrina, \emph{Mathematical Foundations of Quantum Statistical Mechanics} (Kluwer, Dordrecht, 1995).

\bibitem{CGP97}     C. Cercignani, V.I. Gerasimenko and D.Ya. Petrina, \emph{Many- particle Dynamics and Kinetic Equations} (Kluwer Acad. Publ.,
                                                                        Dordrecht, 1997).

\bibitem{Kato}      T. Kato, \emph{Perturbation Theory for Linear Operators} (Springer-Verlag, Berlin, 1995).

\bibitem{GerPolArx} V.I. Gerasimenko and D.O. Polishchuk, arXiv:1001.3893, p.30, 2010.

\bibitem{JStatMech} V.I. Gerasimenko and V.O. Shtyk, J. Stat. Mech., P03007 (2008).

\bibitem{GVI}       V.I. Gerasimenko, Operator Theory: Adv. and Appl. {\bf 191}, 341 (2009). (arXiv:0804.1153, 2008).

\bibitem{GVIa}      V.I. Gerasimenko, arXiv:0908.2797, 2009.



\end{thebibliography}
\end{document}